# The Quantum Model of Spinning Black Holes

V. P.Neznamov [*], S.Yu.Sedov[†], V.E.Shemarulin[‡]

Russian Federal Nuclear Center – All-Russian Research Institute of Experimental Physics

(RFNC-VNIIEF), 37 Mira Ave, Sarov, Nizhny Novgorod region, 607188, Russia

Abstract

We propose a quantum model of spinning black holes with the integrable ring singularities. For the charged Kerr-Newman quantum metric, the complete regularization takes place at fixing of the maximal (cut-off) energy of gravitons, $k_{UV}^{reg} = \hbar c / R_S^{reg}$ .

The domains of existence of one, two and several event horizons $r_q$ are shown depending on the parameters of modified Kerr and Kerr-Newman metrics.

[*] vpneznamov@vniief.ru, vpneznamov@mail.ru
[†] SYuSedov@vniief.ru
[‡] VEShemarulin@vniief.ru

## 1. Introduction

One of the main theme in the gravitation theory is the study of black hole formation in the gravitational collapse of the compact objects. In the black holes theory, there are three main problems which arouse the interest of researchers.

The first problem is connected with the existence of the singularities under the event horizons of the Schwarzschild, Reissner-Nordström, Kerr and Kerr-Newman classical solutions of the general relativity (GR) [1]. The singularities lead to the infinitely large tidal gravitational forces. It breaks a forecasting power of GR.

The second problem is connected with the non-physical quantity in domains under the inner event horizons (see, for example, [2]).

The third problem is connected with the absence of physically significant stationary states of the particle with real energy in the classical black hole fields [3] - [17]. This problem applies to the motion of scalar particles, photons and fermions. In all cases there is a quantum-mechanical mode of particle "fall" on the event horizons [18]. The modes of "a particle in the classical black hole field with the event horizons of zero thickness" are singular. The observed singularities do not allow us to apply quantum mechanics apparatus in full. So, it is necessary to change an initial formulation of the physical problem.

As for regularization of solutions under the black hole horizons in classical version, there are a lot of investigations (see, for example, [19] - [21]). One of the approach to regularization is based on the Markov limiting curvature conjecture [22] which restricts the invariants of GR to finite quantities.

The other tendency of the last decades is the use of the quantum conceptions for the solution of a problem of black hole regularization. Now there are a lot of quantum models of black holes in the literature (see, for example, [23] - [25][§]

In papers [26], [27], the quantum description of black holes is presented for modified Schwarzschild (S) and Reissner-Nordström (RN) geometries. Black holes contain a quantum core described by coherent states of gravitons. Since coherent states cannot contain states with arbitrary short wave lengths of gravitons, classical central singularities of the black holes become integrable singularities with finite tidal forces. In papers [26], [27], the short wave lengths are removed by cut-off of graviton's momenta (energies). As a result, the maximum momentum (maximum energy) of gravitons appears[**]

$$k_{UV} = \frac{\hbar}{R_S}. \qquad (1)$$

---

[§] The order of references [23] - [25] is taken from the paper [26].
[**] We will use units with $c = 1$.

Here, as in [26], [27], for convenience, we introduce a parameter $R_S$ identifiable with a radius of a quantum core. The values $k_{UV}$ and $R_S$ are the free parameters of a theory which definitely connected with each other according to Eq. (1).

In the future quantum theory of gravitation, the energy cut-off will be substituted by strict integration and the absence of short wave lengths of gravitons in coherent states will be a natural result of applying a more perfect quantum theory.

The aim of our effort was to extend the approach of papers [26], [27] to the modified Kerr (K), Kerr-Newman (KN) geometries describing regular quantum spinning black holes either with the single (outer) event horizon or without event horizons. Below for short, the notation "quantum spinning black holes" implies spinning quantum cores either with event horizons or without event horizons.

In sections 2 - 5, we propose a regular quantum model of spinning black holes. In section 6, we analyze the parameters of our model, at which there are not the inner event horizons. In section 7, we discuss the obtained results. In section 8, the brief conclusion is provided.

## 2. Kerr and Kerr-Newman quantum space-time

We will use the extension of the classical Kerr metric (Cürses-Cürsey metric [28])

$$ds^2 = \left(1 - \frac{2r\,m(r)}{\rho^2}\right)dt^2 + \frac{4a\,r\,m(r)\sin^2\theta}{\rho^2}dt\,d\varphi - \frac{\rho^2}{\Delta}dr^2 - \rho^2 d\theta^2 - \frac{\Sigma \sin^2\theta}{\rho^2}d\varphi^2, \qquad (2)$$

where $m(r)$ is the mass function

$$\rho^2 = r^2 + a^2 \cos^2\theta, \qquad (3)$$

$$\Delta = r^2 - 2r\,m(r) + a^2, \qquad (4)$$

$$\Sigma = \left(r^2 + a^2\right)^2 - a^2 \Delta \sin^2\theta, \qquad (5)$$

$$a = \frac{J}{M}. \qquad (6)$$

The classical Kerr metric describes the geometry of an uncharged $(Q=0)$ spinning black hole with a point mass $M$ and an angular momentum $J$. The representation (2) of the classical Kerr metrics is convenient for us to construct quantum model of spinning black holes.

The components of the Einstein tensor for metric (2) can be presented as those in paper [29]:

$$G_0^{\;0} = 2\frac{r^4 + \left(\rho^2 - r^2\right)^2 + a^2\left(2r^2 - \rho^2\right)}{\rho^6}m' - \frac{r a^2 \sin^2\theta}{\rho^4}m'', \qquad (7)$$

$$G_1^{\ 1} = 2\frac{r^2}{\rho^4}m', \tag{8}$$

$$G_2^{\ 2} = 2\frac{\rho^2 - r^2}{\rho^4}m' + \frac{r}{\rho^2}m'', \tag{9}$$

$$G_3^{\ 3} = 2\frac{2r^2(\rho^2 - r^2) + a^2(\rho^2 - 2r^2)}{\rho^6}m' + \frac{r(a^2 + r^2)}{\rho^4}m'', \tag{10}$$

$$G_0^{\ 3} = 2\frac{a(2r^2 - \rho^2)}{\rho^6}m' - \frac{ar}{\rho^4}m''. \tag{11}$$

Here, $m' = \dfrac{dm}{dr}$, $m'' = \dfrac{d^2 m}{dr^2}$.

The important observation (see paper [29]) is the fact that the Einstein tensor is linear in derivatives of a mass function $m(r)$. Any linear decomposition of the mass function

$$m = m_1(r) + m_2(r) \tag{12}$$

leads to the linear decomposition of the Einstein tensor

$$G_\beta^{\ \alpha}(m, a) = G_\beta^{\ \alpha}(m_1, a) + G_\beta^{\ \alpha}(m_2, a). \tag{13}$$

In case of $J = 0, Q = 0$, metric (2) turns into a classical metric S with

$$m_S = GM. \tag{14}$$

In case of $J = 0, Q \neq 0$, metric (2) turns into a classical metric RN with

$$m_{RN}(r) = GM - \frac{GQ^2}{2r}. \tag{15}$$

It is important that for the classical K and KN metrics, the mass functions are not changed

$$m_K = m_S = GM, \tag{16}$$

$$m_{KN}(r) = m_{RN}(r) = GM - \frac{GQ^2}{2r}. \tag{17}$$

By substitution of (16), (17) into (2), we obtain classical K and KN metrics in Boyer-Lindquist coordinates [30]. Hence, it follows that the rotation does not change mass functions.

The second property of the mass functions $m(r)$ is theirs spherical symmetry. The mass functions do not depend on $\theta, \varphi$ angles.

The above properties of the mass functions allow us to fully use the mathematical apparatus, previously used in [26], [27] for quantum metrics S and RN. Let us note that in papers [26], [27], the classical potentials are

$$V_S(r) = \frac{m_S}{r}, \tag{18}$$

$$V_{RN}(r) = \frac{m_{RN}}{r}. \tag{19}$$

Here, in accordance with (14) and (15) $m_S = GM$, $m_{RN}(r) = GM - \frac{GQ^2}{r}$.

Since the mass functions $m(r)$ do not depend on rotation and are spherical symmetric, so the average values of the quantum fields $\Phi$ reproducing the classical potentials $V_S(r)$ and $V_{RN}(r)$ are identical for metrics with rotation and without rotation.

Then, for the quantum case, metric (2) is corrected by the following substitutions.

The Kq metric $(Q=0)$:

$$m_{Kq}(r) = m_{Sq}(r) = GM \frac{2}{\pi} \text{Si}\left(\frac{r}{R_S}\right). \tag{20}$$

The KNq metric $(Q \neq 0)$:

$$m_{KNq} = m_{RNq} = GM \frac{2}{\pi} \text{Si}\left(\frac{r}{R_S}\right) - \frac{GQ^2}{2r}\left(1 - \cos\left(\frac{r}{R_S}\right)\right). \tag{21}$$

In (20) and (21) $m_{Kq}(r), m_{Sq}(r), m_{KNq}(r), m_{RNq}(r)$ are the quantum mass functions for the modified Kerr, Schwarzschild and Kerr-Newman, Reissner-Nordström geometries; $\text{Si}\left(\frac{r}{R_S}\right) = \int_0^{r/R_S} \frac{\sin x}{x} dx$ is an integral sine.

### 3. Effective energy-momentum tensor

Since quantum metric (2) with mass functions (20), (21) is no longer a vacuum solution of GR equations, there must exist non-zero diagonal components of the energy-momentum tensor $T_\mu^{\ \nu} = \frac{G_\mu^{\ \nu}}{8\pi G}$:

$$T_0^{\ 0} = \rho_\varepsilon, \ T_1^{\ 1} = -p_r, \ T_2^{\ 2} = -p_\theta, \ T_3^{\ 3} = -p_\varphi. \tag{22}$$

Here, $\rho_\varepsilon$ is an energy density, $p_r$ is a pressure, $p_\theta$, $p_\varphi$ are the tensions.

For a quantum KN metric, the total energy determined by the volume integral of energy density $\rho_\varepsilon(r, \theta)$ is

$$E = \int T_0^{\,0} \sqrt{-g}\, dV = \frac{1}{4G} \int_0^\infty dr \int_{-1}^{1} d\mu \left(r^2 + a^2 \mu^2\right) G_0^{\,0}(r,\mu) =$$

$$= \frac{1}{4G} \int_0^\infty dr \int_{-1}^{1} d\mu \left[ 2\frac{r^4 + \left(\rho^2 - r^2\right)^2 + a^2\left(2r^2 - \rho^2\right)}{\rho^4} m'_{KN_q} - \frac{r a^2\left(1-\mu^2\right)}{\rho^2} m''_{KN_q} \right] = \quad (23)$$

$$= \frac{1}{4G} \int_0^\infty dr \left\{ \left[ 8 - 4\frac{r}{a}\operatorname{arctg}\frac{a}{r} \right] m'_{KN_q} + \left[ 2r - 2\frac{r^2}{a}\operatorname{arctg}\frac{a}{r} - 2a \operatorname{arctg}\frac{a}{r} \right] m''_{KN_q} \right\} =$$

$$= M + \frac{1}{2}\frac{|J|}{R_S} - \frac{\pi}{16}\frac{Q^2 |J|}{M}\frac{1}{R_S^2} = M + \frac{1}{2}\frac{|J|}{\hbar} k_{UV} - \frac{\pi}{16}\frac{Q^2 |J|}{M}\frac{k_{UV}^2}{\hbar^2}.$$

In (23), for metric (2), $\sqrt{-g} = \rho^2 \sin\theta = \left(r^2 + a^2\mu^2\right)\sin\theta$, $\mu = \cos\theta$.

The last equality in (23) was obtained by using the relation of $R_S = \hbar/k_{UV}$, where $k_{UV}$ is the maximum (cut-off) energy of a graviton.

The total energy $E$ is finite. For K $(Q=0)$ and KN $(Q \neq 0)$ metrics in the last equality of the expression (23), there is the summand proportional to an angular momentum modulus $|J|$ and to maximum (cut-off) energy of graviton $k_{UV}$. For KN $(Q \neq 0)$ metric, in (23), there is also a summand $\sim \dfrac{Q^2 |J|}{M} k_{UV}^2$.

In the absence of cut-off $(k_{UV} \to \infty)$, the total energy of a spinning black hole would be infinite, obviously non-physical value.

In the theory under consideration, a radius $R_S$ is a free parameter.

For quantum S [26] and RN [27] metrics, the total energies of black holes are independent of $R_S$ and are equal to $E = M$.

For the quantum KN metric, we can obtain the similar result if we assume that radius $R_S$ is equal to

$$R_S^{reg} = \frac{\pi}{8}\frac{Q^2}{M}. \quad (24)$$

Expression (24) is close to a classical radius of a charged particle $R^{cl} = \dfrac{Q^2}{M}$.

It is important that at $R_S = R_S^{reg}$, the main physical characteristics of the quantum black hole with the Kerr-Newman geometry become regular.

Let us consider the behavior of the quantum mass function $m_{KNq}(r)$ (see (21)) and its first and second derivatives in the neighbourhood of $r = 0$. These dependencies are given in Table 1.

Table 1. Dependencies $m_{KNq}(r), m'_{KNq}(r), m''_{KNq}(r)$ at $r \to 0$.

|  | $R_S \neq R_S^{reg}$ | $R_S = R_S^{reg}$ |
|---|---|---|
| $m_{KNq}\big|_{r \to 0}$ | $\left(GM\dfrac{2}{\pi} - \dfrac{GQ^2}{4R_S}\right)\left(\dfrac{r}{R_S}\right)$ | $\dfrac{1}{18}\dfrac{GM}{\pi}\left(\dfrac{r}{R_S^{reg}}\right)^3$ |
| $m'_{KNq}\big|_{r \to 0}$ | $\left(GM\dfrac{2}{\pi} - \dfrac{GQ^2}{4R_S}\right)\left(\dfrac{1}{R_S}\right)$ | $\dfrac{1}{6}\dfrac{GM}{\pi R_S^{reg}}\left(\dfrac{r}{R_S^{reg}}\right)^2$ |
| $m''_{KNq}\big|_{r \to 0}$ | $\left(-\dfrac{2}{3}\dfrac{GM}{\pi R_S^2} + \dfrac{GQ^2}{8R_S^3}\right)\left(\dfrac{r}{R_S}\right)$ | $\dfrac{1}{3}\dfrac{GM}{\pi\left(R_S^{reg}\right)^2}\left(\dfrac{r}{R_S^{reg}}\right)$ |

We see that at $R_S \neq R_S^{reg}$, the mass function $m_{KNq}(r) \sim r$, while at $R_S = R_S^{reg}$ $m_{KNq}(r) \sim r^3$. In the first case, the singularities of the components $T_\mu^{\ \nu}$ are integrable. In the second case, the components $T_\mu^{\ \nu}$ are nonsingular.

Really, at $R_S = R_S^{reg}$, the diagonal components of the energy-momentum tensor, determined by formula (22) at $r \to 0$, are equal to

$$T_0^{\ 0}\big|_{r \to 0} = \begin{cases} \sim \left(-r^2/\mu^4\right) & \text{at } \mu \neq 0, \pm 1, \\ \text{const} > 0 & \text{at } \mu = 0, \\ \sim r^4 & \text{at } \mu = \pm 1, \end{cases} \qquad T_1^{\ 1}\big|_{r \to 0} = \begin{cases} \sim r^4/\mu^4 & \text{at } \mu \neq 0, \\ \text{const } 1 > 0 & \text{at } \mu = 0, \end{cases}$$

$$T_2^{\ 2}\big|_{r \to 0} = \begin{cases} \sim r^2/\mu^2 & \text{at } \mu \neq 0, \\ \text{const } 2 > 0 & \text{at } \mu = 0, \end{cases} \qquad T_3^{\ 3}\big|_{r \to 0} = \begin{cases} \sim r^2/\mu^4 & \text{at } \mu \neq 0, \\ \text{const } 3 & \text{at } \mu = 0. \end{cases} \qquad (25)$$

In the Boyer-Lindquist coordinates, the point of $r = 0, \mu = 0$ is a ring singularity of the classical KN metric. For all the components $T_\mu^{\ \nu}$, the integration over volume in the domain of $r = 0$ leads to finite values.

### 4. Kretschmann scalar

Let us consider the behavior of the Kretschmann scalar at $r \to 0$. For brevity, let us introduce the denomination of $K = R^{\alpha\beta\gamma\delta} R_{\alpha\beta\gamma\delta}$.

At $R_S = R_S^{reg}$,

$$K\big|_{r \to 0} = \begin{cases} \sim r^4/\mu^4 & \text{at } \mu \neq 0, \\ \text{const}_K > 0 & \text{at } \mu = 0. \end{cases} \qquad (26)$$

For the classical KN metric,

$$K_{cl}\big|_{r \to 0} = \begin{cases} \sim 1/\mu^8 & \text{at } \mu \neq 0, \\ \sim 1/r^8 & \text{at } \mu = 0. \end{cases} \qquad (27)$$

For the quantum KN metric with $R_S \neq R_S^{reg}$,

$$K_q\big|_{r \to 0} = \begin{cases} \sim 1/\mu^4 & \text{at } \mu \neq 0, \\ \sim 1/r^4 & \text{at } \mu = 0. \end{cases} \tag{28}$$

Thus, the fixation of $R_S = R_S^{reg}$ leads, in the neighborhood of $r = 0$, to finite values of the Kretschmann scalar, including the domain of ring singularity of the classical KN metric of $\mu = 0, r = 0$.

## 5. Space-time structure of the Kerr and Kerr-Newmann geometries

Let us consider the metric of the classical and quantum K and KN geometries in the Kerr-Schild coordinates

$$ds^2 = dt^2 - dx^2 - dy^2 - dz^2 - \left(r^4 + a^2 z^2\right)^{-1} 2r^3 m(r) \times$$
$$\times \left\{ \left(r^2 + a^2\right)^{-1} \left[ r(xdx + ydy) - a(xdy - ydx) \right] + r^{-1} zdz + dt \right\}^2 \tag{29}$$

For the classical KN metric at $r = R^{cl}/2 = Q^2/2M$, $m_{KN} = 0$ and above metric becomes flat. In this case, the outer KN solution is joined with the inner flat space-time metric (see, for example, [31]). For the quantum KN metrics at $R_S = R_S^{reg}$, the mass function $m_{KNq}(r)$ is not zero in the range $r \in (0, \infty)$. I. e., for the quantum KN metrics in the range $r \in (0, \infty)$, there is a curved space-time everywhere.

## 6. Event horizons of spinning quantum black holes

For the KN metric, let us introduce, in formula (4), the denotation of

$$\Delta = r^2 f_{KN}. \tag{30}$$

For the classical KN metric, $m_{KN}(r) = GM - \dfrac{GQ^2}{2r}$.

Then,

$$f_{KN} = 1 - \frac{2GM}{r} + \frac{GQ^2}{r^2} + \frac{a^2}{r^2}. \tag{31}$$

Equality $f_{KN} = 0$ determines the radii of outer and inner event horizons

$$R_\pm = GM \pm \sqrt{G^2 M^2 - GQ^2 - a^2}. \tag{32}$$

If $GM^2 > Q^2 + a^2$, then,

$$f_{KN} = \left(1 - \frac{R_+}{r}\right)\left(1 - \frac{R_-}{r}\right). \tag{33}$$

The case of $GM^2 = Q^2 + a^2$ corresponds to the extreme metric with the single event horizon:

$$R_{\pm} = GM. \tag{34}$$

The case of $GM^2 < Q^2 + a^2$ corresponds to the naked singularity without event horizons. In this case, $f_{KN} > 0$.

For the quantum KN metric, a mass function $m_{KNq}$ is provided in (21). Then,

$$f_{KNq} = 1 - \frac{R_H}{r}\frac{2}{\pi}\text{Si}\left(\frac{r}{R_S}\right) + \frac{R_Q^2}{r^2}\left(1 - \cos\left(\frac{r}{R_S}\right)\right) + \frac{a^2}{r^2}. \tag{35}$$

In (35), $R_H = 2GM$; $R_Q^2 = GQ^2$.

Equality $f_{KNq} = 0$ determines possible event horizons $r_q$:

$$1 - \frac{R_H}{r_q}\frac{2}{\pi}\text{Si}\left(\frac{r_q}{R_H}\frac{1}{R_S/R_H}\right) + \beta_1\frac{R_H^2}{4r_q^2}\left(1 - \cos\left(\frac{r_q}{R_H}\frac{1}{R_S/R_H}\right)\right) + \beta_2\frac{R_H^2}{4r_q^2} = 0. \tag{36}$$

Here, $\beta_1 = \frac{R_Q^2}{R_H^2/4}, \beta_2 = \frac{a^2}{R_H^2/4}$; for the S metric, $\beta_1 = \beta_2 = 0$; for the RN metric $\beta_1 \neq 0, \beta_2 = 0$; for the K metric $\beta_1 = 0, \beta_2 \neq 0$; $\beta_1 + \beta_2 = 1$ is the analog of the extreme KN metric; $(\beta_1 + \beta_2) > 1$ is the parametric analog of the naked KN singularity[††].

For the fixed values of $\beta_1, \beta_2$, the equation (36) defines the curves on the plane $\left(\frac{r_q}{R_H}, \frac{R_S}{R_H}\right)$ implicitly:

$$\varphi_{\beta_1,\beta_2}\left(\frac{r_q}{R_H}, \frac{R_S}{R_H}\right) = 0. \tag{37}$$

The curves (37) for some values of $\beta_1, \beta_2$ determined by using Maple software package are presented in Fig. 1, Fig. 2.

---

[††] For the quantum KN black hole, there is no notion of naked singularity. In this case, in the center of the system, there is a quantum core with possible event horizons (see Fig. 2).

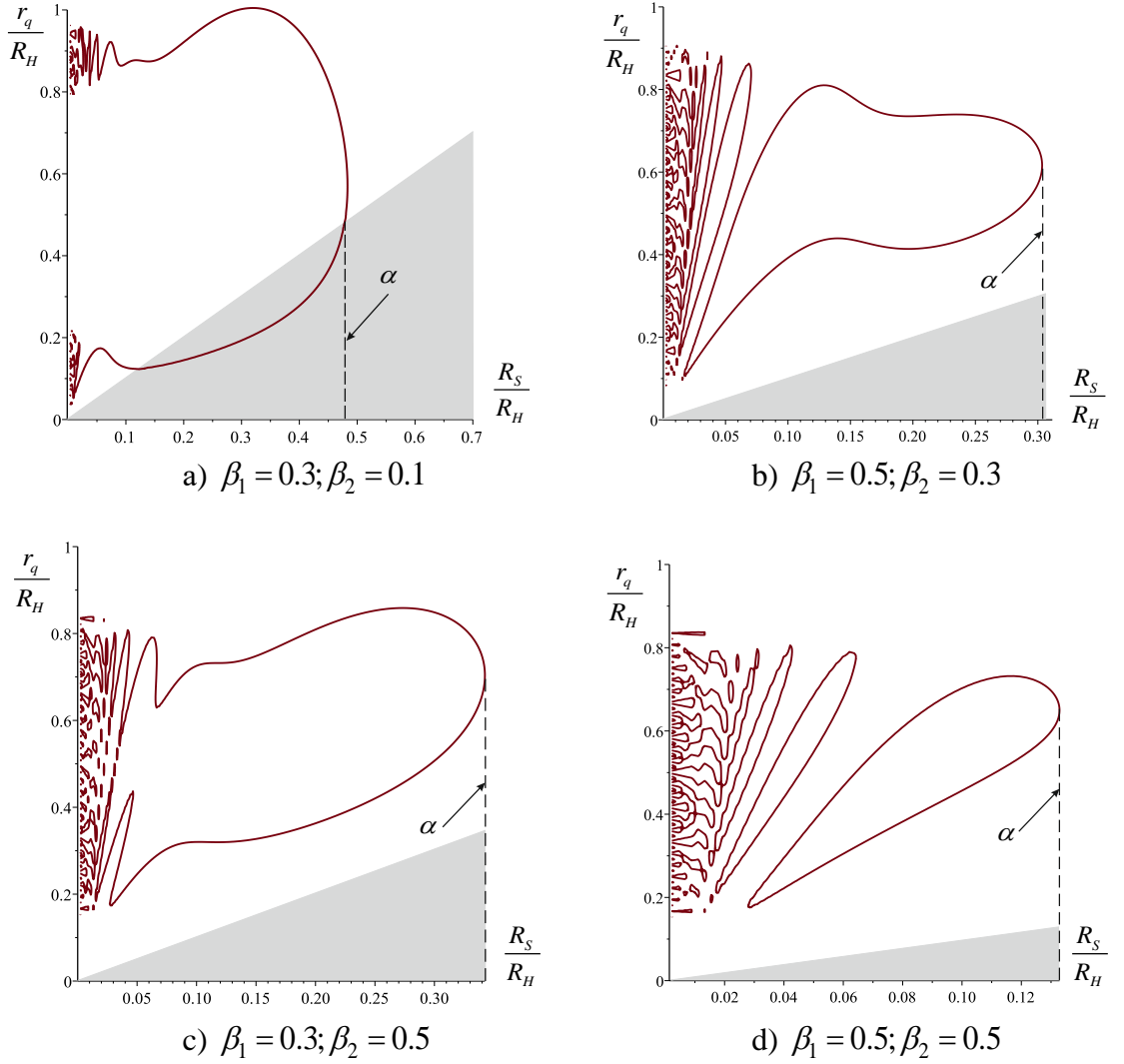

Fig. 1. Radii of quantum event horizons $r_q$ for different values of $R_S$ and $(\beta_1 + \beta_2) \leq 1$.

It is seen from the Fig. 1 that at small values $R_S/R_H$, multitude event horizons are available. With increase in $R_S$, the number of horizons decreases down to one $(r_q)_+$. Further increase in $R_S$ leads to disappearance of all the event horizons.

As against the classical extreme RN black hole with a single event horizon $R_\pm = R_H$ for the quantum RN black hole with $\beta_1 + \beta_2 = 1$, the pattern described for the values of $0 < (\beta_1 + \beta_2) < 1$ is reproduced qualitatively. In this case, the disappearance of event horizons occurs at values of $R_S/R_H > 0.134$ (see Fig. 1d).

The shadowy areas in Fig. 1 correspond to the parameters of $r_q < R_S$. Such parameters cannot be parameters of black holes.

For classical RN black holes, the case of $(\beta_1+\beta_2)>1$ corresponds to the naked RN singularity. For quantum RN black holes at $(\beta_1+\beta_2)>1$ and $R_S/R_H<0.1$, there are several event horizons (see Fig. 2). With increase in $R_S$, the event horizons disappear.

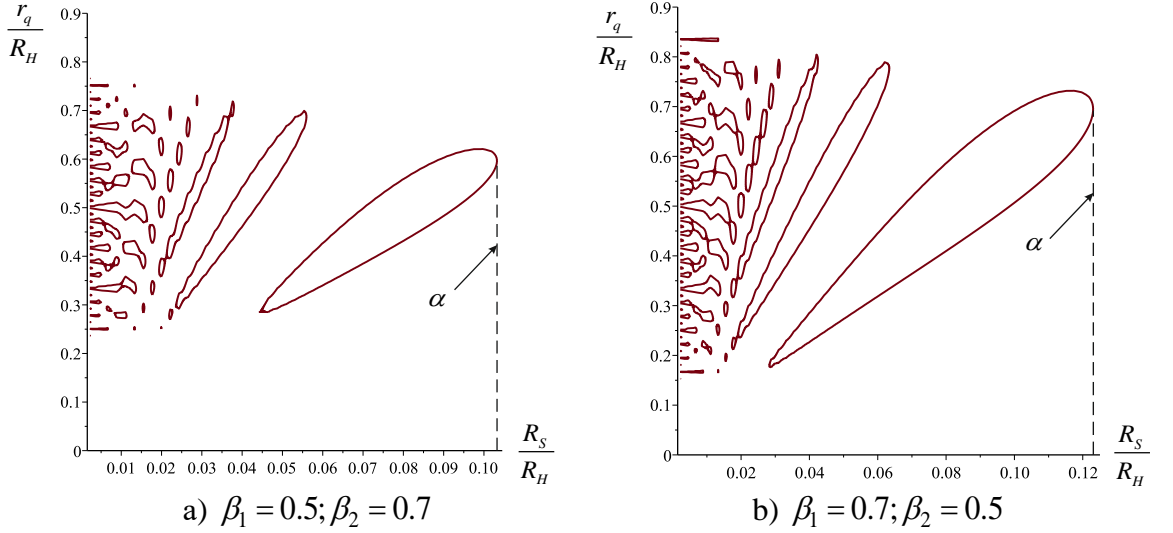

a) $\beta_1=0.5; \beta_2=0.7$      b) $\beta_1=0.7; \beta_2=0.5$

Fig. 2. Radii of quantum event horizons $r_q$ for different $R_S$ and $(\beta_1+\beta_2)>1$ values.

It is well known that the classical Kerr and Kerr-Newman geometries are non-physical in the domain under the inner event horizon. In the paper [32], the authors suppose the existence of a dependence the angular momentum on the radius $(|J|\sim a\sim r)$ for absence of the inner event horizons in the spinning black holes. In our model, if there is a free parameter $R_S$, we always can choose a ratio $R_S/R_H \geq \alpha(\beta_1,\beta_2)$ which provides either an absence of the event horizons or an existence of the single event horizon $(r_q)_+$.

So, for Fig. 1a $\alpha \geq 0.485$, for Fig. 1b $\alpha \geq 0.306$, for Fig. 1c $\alpha \geq 0.345$, for Fig. 1d $\alpha \geq 0.134$, for Fig. 2a $\alpha \geq 0.104$, for Fig. 2b $\alpha \geq 0.125$.

For the regular quantum Kerr-Newman geometry with $R_S = R_S^{reg}$, the ratio $\dfrac{R_S^{reg}}{R_H}$ is equal to $\dfrac{\pi}{16}\dfrac{Q^2}{M^2 G}$. The expression $\left(\dfrac{R_S}{R_H}\right)_1 \geq \alpha(\beta_1,\beta_2)$ adds up to the fulfilment of the next relation between a charge and mass of quantum black hole

$$\frac{\pi}{16}\frac{Q^2}{M^2 G} \geq \alpha(\beta_1,\beta_2).$$

Furthermore, note that at $R_S = R_S^{reg}$ and $a^2 \gg R_Q^2$, the quantum KN black hole does not have event horizons.

## 7. Discussions of the results

In our paper we propose a quantum model of spinning black holes. The model is a result of the extension of the approach of papers [26], [27] to Kerr and Kerr-Newman geometries.

In papers [26], [27], the phenomenological description of quantum black holes is provided for modified S and RN geometries. The quantum black holes are described by coherent states of gravitons with the maximum (cut-off) energy of $k_{UV} = \hbar/R_S$.

For classical and quantum K and KN metrics, we use metric (2) with a mass function $m(r)$. The classical S and K metrics and the classical RN and KN metrics are described by identical functions, respectively.

$$m_S = m_K = GM, \tag{38}$$

$$m_{RN} = m_{KN} = GM - \frac{GQ^2}{2r}. \tag{39}$$

For quantum metrics with rotation we use the similar equalities

$$m_{Sq} = m_{Kq} = GM \frac{2}{\pi} \text{Si}\left(\frac{r}{R_S}\right), \tag{40}$$

$$m_{RNq} = m_{KNq} = GM \frac{2}{\pi} \text{Si}\left(\frac{r}{R_S}\right) - \frac{GQ^2}{2r}\left(1 - \cos\left(\frac{r}{R_S}\right)\right). \tag{41}$$

Functions $m_{Sq}(r)$ and $m_{RNq}(r)$ were used in papers [26], [27] for the modified quantum S and RN metrics.

An important feature of a mass function $m(r)$ is spherical symmetry allowing the use of approaches of papers [26], [27] for regularization of spinning K and KN black holes.

In the proposed model with modified K and KN metrics, the divergence in the domain of ring singularity $(\mu = 0, r = 0)$ becomes integrable. For example, the Kretschmann scalar for classical K and KN metrics is $R_{\alpha\beta\gamma\delta}R^{\alpha\beta\gamma\delta} \sim 1/r^8$. In our model for quantum K and KN metrics, $R_{\alpha\beta\gamma\delta}R^{\alpha\beta\gamma\delta} \sim 1/r^4$.

The total energy for quantum K and KN metrics is finite and equal to:

$$E = \int T_0^{\ 0} \sqrt{-g}\, dV = M + \frac{1}{2}\frac{|J|}{\hbar}k_{UV} - \frac{\pi}{16}\frac{Q^2|J|}{M}\frac{k_{UV}^2}{\hbar^2}. \tag{42}$$

For a quantum K $(Q=0)$ metric, the second summand is proportional to the maximum (cut-off) energy of gravitons. In case of quantum KN $(Q \neq 0)$ metric, the additional third summand is proportional to the square of the maximum (cut-off) energy of gravitons.

For the quantum KN metric, the situation for regularized black holes essentially changes if we assume that free parameter of the theory $R_S$ is equal to $R_S^{reg} = \frac{\pi}{8}\frac{Q^2}{M}$.

At $R_S = R_S^{reg}$, the Kretschmann scalar in the domain $(\mu=0, r=0)$ is equal to $\text{const}_K > 0$. All the components of energy-momentum tensor $T_\mu^{\ \nu}$ are finite at $r \to 0$. For all components $T_\mu^{\ \nu}$, the integration over volume in the domain $r=0$ leads to the finite values. The total energy is $E = M$. The same value for energy $E$ in [26], [27] was obtained for the S and RN metrics[‡‡].

## 8. Conclusions

On the basis of papers [26], [27], the regular quantum model of spinning black holes is proposed. For the charged quantum KN metric, when fixing the maximum (cut-off) energy of a graviton ($k_{UV}^{reg} = \hbar/R_S^{reg}$, where $R_S^{reg} = \frac{\pi}{8}\frac{Q^2}{M}$), the complete regularization of the metric is demonstrated in the domain of ring singularity $(\mu=0, r=0)$.

On the plane of $\left(\frac{r_q}{R_H}, \frac{R_S}{R_H}\right)$, the domains of existence of one, two or several event horizons are shown depending on the parameters of the modified K and KN metrics.

The absence of the inner event horizons is achieved by a choice of the free parameter $R_S$ satisfying the relation

$$\frac{R_S}{R_H} \geq \alpha(\beta_1, \beta_2). \qquad (43)$$

If $\frac{R_S}{R_H} = \alpha(\beta_1, \beta_2)$, then it corresponds to an existence of only single (outer) event horizon.

If $R_S = R_S^{reg}$, then the expression (43) adds up to the fulfillment of the relation between charge and mass of quantum black hole

$$\frac{\pi}{16}\frac{Q^2}{M^2 G} \geq \alpha(\beta_1, \beta_2).$$

---

[‡‡] As $R_S = R_S^{reg}$, it is easy to prove the compatibility of the Maxwell equations with the components of the Einstein tensor for the charged quantum RN and KN metrics. The proof will be give in our next paper "Towards to a quantum model of an electron with the zero self-energy".

## Acknowledgements

The research was carried out within the framework of the scientific program of the National Center for Physics and Mathematics, the project "Particle Physics and Cosmology".

The authors thank A.L.Novoselova for the essential technical assistance in preparation of the paper.